\def\figsize{8.7cm}

\def\rn{}
\def\nn#1 #2{#2. #1}				
\def\nnn#1 #2 #3{#2. #3. #1}			
\def\nnnn#1 #2 #3 #4{#2. #3. #4 #1}		
\def\nnnnn#1 #2 #3 #4 #5{#2. #3. #4 #5. #1}	
\def\dualand{ and\hbox{ }}				
\def\multiand{, and\hbox{ }}				
\def\rf#1;#2;#3;#4;#5 {{\frenchspacing\par\rn#1, #3 {\bf #4}, #5 (#2). \par}}
\def\rg#1;#2;#3;#4;#5;#6 {{\frenchspacing\par\rn#1, #3 {\bf #4}, #5 (#2). \par}}
\def\rfproc#1;#2;#3;#4;#5;#6 {{\frenchspacing\par\rn#1 #2, in {\it #3}, ed. #4 (#5: #6)\par}}
\def\rfprocp#1;#2;#3;#4;#5;#6;#7 {{\frenchspacing\par\rn#1 #2, in {\it #3}, ed. #4 (#5: #6), p#7\par}}

\def\rg#1;#2;#3;#4;#5;#6 {\par\rn#1 #2, {\it #3}, {\bf #4}, #5 (``#6'') \par}
\def\rf#1;#2;#3;#4;#5 {\par\rn#1, {\it #3}, {\bf #4}, #5 (#2)\par}
\def\rfbook#1;#2;#3;#4;#5 {{\frenchspacing\par\rn#1, {\it #3} (#4: #5, #2)\par}}
\def\rfproc#1;#2;#3;#4;#5;#6 {{\frenchspacing\par\rn#1 #2, in {\it #3}, ed. #4 (#5: #6)\par}}
\def\rfprocp#1;#2;#3;#4;#5;#6;#7 {{\frenchspacing\par\rn#1 #2, in {\it #3}, ed. #4 (#5: #6), p#7\par}}
\def\rfprep#1;#2;#3 {{\par\frenchspacing\rn#1, #3 (#2)\par}}
\def\rfprepp#1;#2;#3 {{\par\rn#1 #2, #3\par}}

\def\etal{{\frenchspacing\it et al.}}
\def\ie{{\frenchspacing\it i.e.}}
\def\eg{{\frenchspacing\it e.g.}}
\def\etc{{\frenchspacing\it etc.}}

\def\beq#1{\begin{equation}\label{#1}}
\def\eeq{\end{equation}}
\def\beqa#1{\begin{eqnarray}\label{#1}}
\def\eeqa{\end{eqnarray}}

\def\xor{\mathop\mathrm{xor}}

\def\fig#1{Figure~\ref{#1}}
\def\Fig#1{Figure~\ref{#1}}

\def\Sec#1{Section~\ref{#1}}
\def\Sec#1{Section~\ref{#1}}

\def\spose#1{\hbox to 0pt{#1\hss}}
\def\simlt{\mathrel{\spose{\lower 3pt\hbox{$\mathchar"218$}}
     \raise 2.0pt\hbox{$\mathchar"13C$}}}
\def\simgt{\mathrel{\spose{\lower 3pt\hbox{$\mathchar"218$}}
     \raise 2.0pt\hbox{$\mathchar"13E$}}}
\def\simpropto{\mathrel{\spose{\lower 3pt\hbox{$\mathchar"218$}}
     \raise 2.0pt\hbox{$\propto$}}}

\def\ed{\end{document}}

\def\beq#1{\begin{equation}\label{#1}}
\def\eeq{\end{equation}}
\def\beqa#1{\begin{eqnarray}\label{#1}}
\def\eeqa{\end{eqnarray}}

\documentclass[twocolumn,amsmath,nofootinbib]{revtex4} 
\usepackage{amsfonts,amsbsy,epsf} 
\begin{document}






\date{Submitted to {\it MNRAS} October 12 2009, accepted August 22 2010}

\title{Solving the Corner-Turning Problem for Large Interferometers}

\author{Andrew Lutomirski}
\address{Center for Theoretical Physics, Massachusetts Institute of Technology, Cambridge, MA 02139}

\author{Max Tegmark}
\address{Dept.~of Physics \& MIT Kavli Institute, Massachusetts Institute of Technology, Cambridge, MA 02139}

\author{Nevada Sanchez}
\address{Dept.~of Physics \& MIT Kavli Institute, Massachusetts Institute of Technology, Cambridge, MA 02139}

\author{Leo Stein}
\address{Dept.~of Physics \& MIT Kavli Institute, Massachusetts Institute of Technology, Cambridge, MA 02139}

\author{W.~Lynn Urry}
\address{Dept.~of Astronomy, University of California, Berkeley, CA 94720}

\author{Matias Zaldarriaga}
\address{Institute for Advanced Study, Einstein Drive, Princeton, NJ 08540, USA}

\begin{abstract}
The so-called corner turning problem is a major bottleneck for radio telescopes with large numbers of antennas.
The problem is essentially that of rapidly transposing a matrix that is too large to store on one single device; 
in radio interferometry, it occurs because data from each antenna needs to be routed to an array of processors that 
will each handle a limited portion of the data (a frequency range, say) but requires input from each antenna. 
We present a low-cost solution allowing the correlator to transpose its data in real time, without contending for bandwidth, via a butterfly network
requiring neither additional RAM memory nor expensive general-purpose switching hardware.
We discuss possible implementations of this using FPGA, CMOS, analog logic and optical technology, 
and conclude that the corner turner cost can be small even for upcoming massive radio arrays.
\end{abstract}

\keywords{large-scale structure of universe 
--- galaxies: statistics 
--- methods: data analysis}

\pacs{98.80.Es}
  
\maketitle



\setcounter{footnote}{0}

\def\thetamin{\theta_{\rm min}}
\def\thetamax{\theta_{\rm max}}

\section{Introduction}
\label{IntroSec}

There is now strong community interest in building more sensitive radio telescopes, stemming from diverse 
science opportunities that range from planets to pulsars, from black holes to cosmology 
\cite{SKAscienceBook,Jester09,decadal1,decadal2,decadal3,decadal4}. 
However, greater sensitivity requires greater collecting area, which in turn increases cost. 
For large collecting area, interferometers become more cost-effective than single-dish radio telescopes, but pose 
interesting engineering challenges. 

One such challenge is that  for a general large interferometer, the cost grows
quadratically with area, because for an array of $N$ antennas, all $N(N-1)/2\simpropto N^2$ pairs 
of antennas need to be correlated to calculate the so-called visibilities. 
The computational costs thus scale as $N^2$ and dwarf all other costs for large enough $N$.
Fortunately, there are attractive design approaches such as tiling, 
the MOFF-correlator \cite{MoralesMoff} and the omniscope \cite{fftt,omniscopes}
which enable more competitive cost scaling, some having costs that grow as slowly as $N\log N$ with size without losing 
any information \cite{omniscopes}.

\begin{figure}[ht]
\centerline{\epsfxsize=\figsize\epsffile{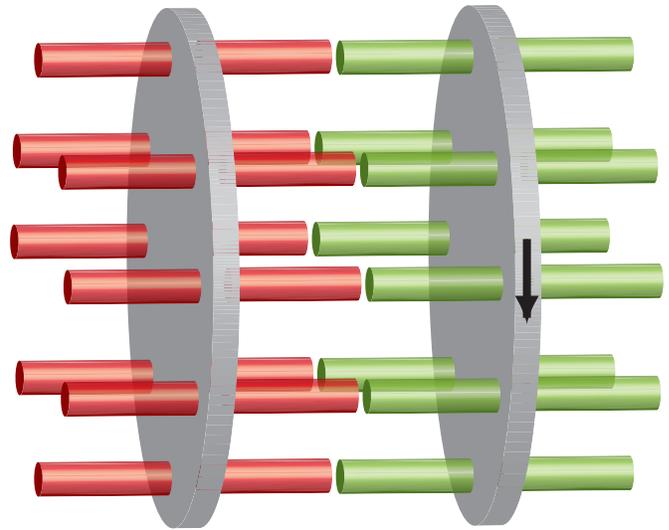}}
\caption{\label{WheelsFig}
Toy example of how the $N=8$ corner turning problem could be solved using moving parts, with 8 devices simultaneously transmitting data into the 8 red/dark grey links 
on the left and 8 other devices receiving data from the 8 green/light grey links on the right. 
After 8 successive $45^\circ$ rotations of the right wheel, all input devices have transmitted the required data to all output devices.
}
\end{figure}

A second bottleneck, common to all the above-mentioned approaches, is what is known as the {\it corner turning problem}.
The problem is essentially that of rapidly transposing a matrix that is too large to store on one single device;
in radio interferometry, it occurs because data from each antenna needs to be routed to an array of processors that 
will each handle a limited portion of the data (a frequency range, say) but requires input from each antenna. 
Most often, the data from each of the $N$ antenna signals can be filtered, amplified, digitized and 
decomposed into channels separately from all other antenna signals.
Suppose that the signal from each antenna is digitized and processed so as to produce a sample stream from each of $M$ separate frequency channels.\footnote{For simplicity, we 
ignore the polarization issue in our discussion, since it can be trivially incorporated by simply doubling $N$ and treating each of the two 
polarization channels from each antenna as an independent data stream.}
If we imagine all this data arranged in an $N\times M$ matrix, then each {\it row} of the matrix thus resides on a separate
physical device (typically an FPGA, a field-programmable gate array).
However, the subsequent computation of UV plane visibilities needs to combine information from all $N$ antennas, separately for each frequency, \ie, 
each {\it column} of the matrix needs to be processed separately and rapidly. 
For a traditional interferometer, this second stage involves multiplying the numbers from the $N(N-1)/2$ pairs of numbers, the so-called ``x-operation'', 
as distinguished from the preceding so-called $f$ operation that Fourier transformed in the time domain. For an omniscope, this 
second stage involves performing a Fourier transform in two or more dimensions \cite{omniscopes}. In either case, the amount of computation of required for the second stage is at least proportional to $N$, so we need $M \gtrsim N$ channels to distribute the computation.
In either case, the corner turning is a major bottleneck, since transferring data between all $~N\times M$ pairs of 
first-stage and second-stage devices requires moving the entire contents of the matrix across a large network. For example, for an $N=64\times 64$ dual polarization 
omniscope sampling at 400 MHz, the corner turner has to route about 13 terabytes per second.
Once this bottleneck has been passed and the second stage has been completed, however, the resulting 
sky maps (or their Fourier transforms) can be time-averaged, dramatically reducing the data rate to 
manageable levels.

Modern radio telescopes have typically adopted one of the following solutions to the corner turning problem:
\begin{enumerate} 
\item Writing the entire matrix to a single, extremely fast, giant memory module where it can be read out transposed, or using some other device with size $O(M^2)$, for example enough wires to turn the corner directly.
\item Routing all the data through an off-the-shelf non-blocking switch.
\item Using enough wires to make all the connections directly.
\end{enumerate}
The first approach has been used by numerous experiments, and the second has been successfully implemented in the packetized CASPER correlator
\cite{CASPER,Parsons08} used by
the PAPER experiment \cite{PAPER}, where $N=32$ (including polarization) is small enough to be handled by a single 10 GB Ethernet switch.  The third is used in some very large correlators such as EVLA and ALMA.
Unfortunately, all of these approaches become expensive for very large $N$, which makes it timely to explore alternative solutions.

Another way of thinking about the corner turning problem is that it involves the only part of an interferometer that 
is not embarrassingly parallel\footnote{Computer scientists say that a problem is ``embarrassingly parallel'' if it can be trivially distributed across a large number of processors that do not need to communicate with each other.}: it is easy to build many antennas, many A/D converters, 
many time-domain Fourier transformers and many correlators acting on separate frequency bands.  
The corner turn is the piece that transposes the data matrix to keep the processing embarrassingly parallel.

The rest of this paper is organized as follows. 
In \Sec{AlgorithmSec}, we present our solution to the corner turning problem, which requires neither general-purpose network switches nor additional memory.
We discuss various physical implementation options in \Sec{ImplementationSec} and summarize our conclusions in \Sec{ConclusionsSec}.

\begin{figure}[t]
\centerline{\epsfxsize=\figsize\epsffile{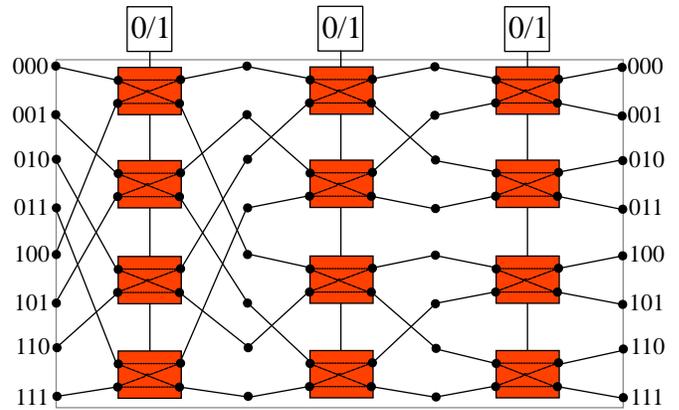}}
\vskip-6.0cm
\caption{\label{ButterflyFig}
How to solve the $N=8$ corner turning problem using a butterfly network, with 8 devices simultaneously transmitting data from the left and 8 
other devices receiving data from the right. 
After the three control bits shown at the top have looped through all 8 combinations 000, 001, 010, 011, 100, 101, 110 and 111, 
all input devices have transmitted the required data to all output devices.
The boxes (``controlled swappers'') have two input wires and either pass them straight through to their output wires
or swap them, depending on whether the control bit (drawn as entering from above) is 0 or 1, respectively.
The 8 inputs are numbered in binary on the left hand side, and we see that
the $1^{\mathrm{st}}$ row of swappers can flip their $1^{\mathrm{st}}$ address bit,
the $2^{\mathrm{nd}}$ row of swappers can flip their $2^{\mathrm{nd}}$ bit, and
the $3^{\mathrm{rd}}$ row of swappers can flip their $3^{\mathrm{rd}}$ bit.
}
\end{figure}

\section{The butterfly algorithm}
\label{AlgorithmSec}


Below we will limit our discussion to the special case where $M=N$, \ie, where the matrix to be transposed is square, or, equivalently, where there are equal numbers of 
devices writing to and reading from the corner turner, since if one has an efficient corner turner for this case, then 
the general case becomes easy to solve.  For example, one can simply pad the matrix with zeros, which is equivalent to inserting dummy data sources or sinks, or combine a few adjacent entries into one larger entry, splitting it at the end if necessary \cite{ATAMemoNonsquare}. We will present an optimized solution for the case where $N$ is a power of 2.


\subsection{The problem}

In the discussion below, we will refer to the concept of a ``link,'' by which we mean some connection between two computers, FPGAs, or any other devices (nodes) that can carry data at a rate of $N$ matrix entries each time a matrix is transposed.  In most cases, this will simply be the rate at which the $f$ stage of the correlator outputs data \emph{for a single antenna}.  In other words, the data rate on a given link is independent of the size of the interferometer.  If the data rate for a single channel exceeds that of the technology we use to build our corner turner (\eg 1 Gbps if we used gigabit Ethernet connections), then we will use bundles of identical connections as our links.

It is easy to solve the corner turning problem with a non-blocking switch that can connect $2N$ links: each source node that starts with one row of the matrix simply transmits all of its data, with each entry in the matrix addressed to the sink node labeled with the column number of that entry.  Each node sends or receives at exactly the link rate, so a general-purpose nonblocking switch can handle all of the data.

Large general-purpose high-speed switches are expensive because they 
are fully non-blocking, allowing any set of input devices to simultaneously transmit to any set of output devices.
This is overkill for the corner turning problem, since we have complete prior knowledge of how data needs to be distributed.
This suggests the possibility of reducing cost by giving up complete generality.

We need each of the $N$ source nodes to transmit data through our corner turner to each of the $N$ sink nodes, with each source node transmitting exactly a $1/N$ fraction of its total data rate to each sink node.  This can be done with a very restricted kind of switch using no memory at all.  Such a switch has $N$ states, selected by a control input $c\in\{0,\ldots,N-1\}$, where the source node labeled with a number $i$ is connected to the sink node $j=p(c,i)$ where the function $p$ has the following properties:
\begin{enumerate}
\item For fixed $i$, all $p(c,i)$ are unique and in the range $0,...,N-1$.
\item For fixed $c$, all $p(c,i)$ are unique and in the range $0,...,N-1$.
\end{enumerate}
In other words, the corner turner performs a different permutation of the inputs at each time step, such that after $N$ steps, 
every input node has been connected to every output node exactly once.

With such a switch, each source node $i$ transmits the $i,j$ entry of the matrix exactly when $p(c,i)=j$, and each sink node $j$ will receive the entire column $j$, albeit in some arbitrary order.  This means that each source node transmits data in a different order, but most receiver or $f$ stage designs should be able to handle this without difficulty.

\subsection{Our solution}

How should we choose the sequence of permutations $p$?
There are clearly vast numbers of permutation sequences that satisfy the two requirements above, since we have $N!$ choices even for $p(0,j)$ alone.

\subsubsection{A mechanical solution}

One simple solution is that defined by the cyclic permutations
\beq{WheelEq}
p(c,i) \equiv c+i\mod N.
\eeq
This choice is illustrated in \fig{WheelsFig} for a toy example where where $N=8$.
If we connect the input devices to the metal bars protruding on the left side and the output devices to the bars protruding to the right, 
then the $N$ successive $45^\circ$ rotations of the right wheel will achieve a complete corner turn where 
every input device has transmitted to every output device.

\subsubsection{The butterfly algorithm}

In practice, one of course needs to accomplish all operations
electronically without large moving parts. 
An elegant method for implementing precisely the cyclic permutations of \fig{WheelsFig}
electronically was discovered about a decade ago by Lynn Urry and implemented 
for the Allen Telescope Array \cite{ATAMemoCornerTurn,ATAMemoCornerTurn2}, but this was unfortunately never 
published in a journal and did not become as widely known as it deserves to be.
The other authors of this paper independently discovered the methods that we will 
describe below, which have the further advantage of being even cheaper to implement.

We schematically illustrate a simple solution in \fig{ButterflyFig},
where the boxes (``controlled swappers'') have two input wires and either pass them straight through to their output wires
or swap them, depending on whether a control bit (drawn as entering from above) is 0 or 1, respectively.
If the $N=8$ inputs $i$ are numbered in binary, then 
the $1^{\mathrm{st}}$ row of swappers can flip their $1^{\mathrm{st}}$ bit,
the $2^{\mathrm{nd}}$ row of swappers can flip their $2^{\mathrm{nd}}$ bit, and
the $3^{\mathrm{rd}}$ row of swappers can flip their $3^{\mathrm{rd}}$ bit.
This means that this corner turner implements the permutations
\beq{ButterflyEq}
p(c,i) \equiv c\xor i,
\eeq
where the integers $c$, $i$ and $j$ are written in binary on the top, left and right sides of \fig{ButterflyFig}, respectively.

This basic network topology where a given node successively ``talks'' with nodes separated by $2^0$, $2^1$, $2^2$, {\etc} appears in
a wide range of electrical engineering and software applications, including the Fast Fourier transform of $N$ numbers, 
and is often referred to as a ``butterfly network''.
When the ``talk'' part is a swapper like in \fig{ButterflyFig}, the resulting network is a special case of a Banyan network \cite{Youssef92} --- a type of general-purpose network switch which is nonblocking for certain permutations (as opposed to fully nonblocking, which would allow nodes to talk to each other in any permutation.)
A key point about the corner turner that we are proposing is that we are {\it not} using it as a general-purpose switch 
but rather with a specific algorithm: to cycle through $N$ very particular permutations, 
which is precisely what is needed to solve the problem at hand.

For comparison, the method in \cite{ATAMemoCornerTurn,ATAMemoCornerTurn2} also uses a Butterfly network, but changes all 
$N\log N$ controlled swappers independently at each control step $c$ instead of using the same setting for 
each of the $\log N$ columns. The latter implementation thus requires $N$ times fewer control input wires, and we will see below how it can be further simplified to cut cost.

The butterfly algorithm we have proposed requires that $N$ be a power of 2. As we will see in \Sec{ImplementationSec},
the cost of a butterfly corner turner is likely to constitute only a small fraction of the total cost of a large $N$ radio array, 
so for a general number of antennas, a one can simply round $N$ up to the nearest power of 
two for the corner turner.



\subsubsection{An even cheaper corner turner using perfect shufflers}

\begin{figure}[ht]
\centerline{\epsfxsize=\figsize\epsffile{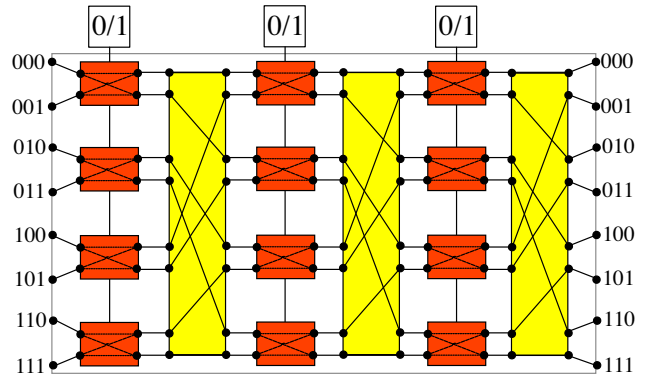}}
\vskip-5.4cm
\caption{\label{ShufflerFig}
An equivalent network to \fig{ButterflyFig}, but reorganized so that the connections after each row 
of swappers are identical. These connections, contained within the tall boxes, are seen 
to correspond to a ``perfect shuffle'', whereby the top half gets interleaved with the bottom half, 
and corresponds to cyclically shifting the address bits that label the inputs on the left-hand-side.
}
\end{figure}

An obvious drawback to the configuration in \Fig{ButterflyFig} is
that the wiring layer between each stage is different, which complicates the manufacturing of a device 
to implement the network.
It turns out that by a suitable permutation of the nodes after each row of swappers,
one can simplify the wiring diagram so that all layers becomes identical, as illustrated in \Fig{ShufflerFig}.

The required permutation performed between the wires after each row of swappers, contained within the tall boxes 
in the figure, turns out to be a so called ``perfect shuffle''
permutation on $N$ elements, which draws its name from card
shuffling. The perfect shuffle (or Faro shuffle) is defined as
\beq{PerfectShuffle}
\left\{
\begin{array}{ll}
j\mapsto 2j          	&\hbox{if}\>\> j<N/2,\\
j\mapsto 2j-N+1 	&\hbox{otherwise},
\end{array}
\right.
\eeq
and corresponds to interleaving the top and bottom halves in the input
\cite{perfect-shuffle}.

\Fig{ShufflerFig} illustrates that if we write the input row $j$ as a binary number composed of $\log_2 N$ bits, 
a perfect shuffle simply permutes its bits cyclically, shifting them all one notch to the left and moving the leftmost bit 
all the way to the right. $\log_2 N$ perfect shuffles thus restores $j$ to its original value.
Since the rows of swappers in \Fig{ShufflerFig} can flip the rightmost bit (exchanging two neighboring rows),
the net effect of the $n^{\mathrm{th}}$ control bit from the left at the top of the figure is thus to control the 
$n^{\mathrm{th}}$ bit from the right of $j$. In other words, \Fig{ShufflerFig} 
corresponds to the same permutation sequence $p(c,i)=c\>\hbox{xor}\>i$ as \Fig{ButterflyFig} except for the trivial 
modification that the control variable $i$ has its bits in reverse order.

To further reduce cost, we can omit the last perfect shuffle layer, since the resulting network is equivalent to the butterfly network with its outputs permuted and retains all of the necessary properties.

\section{Implementation}
\label{ImplementationSec}

\subsection{Layout}

Section~\ref{AlgorithmSec} described the basic layout of the butterfly network.  We can solve the corner turning problem for an $N$-element interferometer using an $N$-link butterfly network.  Since this technique is meant to scale to \emph{large} radio telescopes, an ideal implementation of the butterfly network would be built out of large (but not too large) numbers of identical, inexpensive, and easy-to-connect parts.

In the discussion below, we use $n=\log_2 N$ to refer to the number of layers in the network.

\subsubsection{Network cost}

If we built a butterfly network corner turner out of modular components, then the cost could be roughly computed by counting the number of each type of component.  To build a very large network, or to build many smaller networks, it would be worth the extra effort to design the components so that they would be inexpensive to manufacture and so that as few as possible would be needed.

The simplest set of components to use would be discrete $2\times 2$ switches and single-link cables.  Each controlled swapper layer is $N/2$ identical $2\times 2$ switches arranged in a column and each perfect shuffle layer consists of $N$ cables running between pairs of computers or switches.  For a hypothetical $2^{20} = 1048576$-node interferometer, this comes out to 20 layers, for a total of 10,485,760 switches and 20,971,520 cables.  This is doable (the interferometer would be expensive enough that this would most likely be only a small part of the cost), but connecting 20 layers of over one million cables without making mistakes would be tedious at best.

\subsubsection{How to further reduce the cost}

There is room for a large improvement in the number of parts needed, though: printed circuit boards containing hundreds of components are inexpensive (in 2009, 12 inch by 14 inch circuit boards can be fabricated for less than \$30 each, even in small volumes, assuming that the circuit fits on two layers) and cables are available that can carry many links worth of bandwidth.  (Of course, using large cables may only be useful internally---unless multiple source or sink nodes are on the same board, the input and output links must each be on its own cable.)

To optimize the perfect shuffles, we will use a simple property of a cyclic bit shift: an $n$-bit binary number $b_{n-1}b_{n-2}\ldots b_0$ can be circularly shifted one bit to the left by first circularly shifting the leftmost $n-k$ bits one bit to the left, giving $b_{n-2}\ldots b_{k+1}b_{k}b_{n-1}b_{k-1}\ldots b_0$ and then shifting the rightmost $k+1$ bits one bit to the left, giving $b_{n-2}\ldots b_{k+1}b_{k}b_{k-1}\ldots b_0b_{n-1}$ as desired.  (For example, $n=5$ and $k=2$ starts with $b_4 b_3 b_2 b_1 b_0$, maps it to $b_3 b_2 b_4 b_1 b_0$, and finally maps it to $b_3 b_2 b_1 b_0 b_4$.)

\begin{figure}[ht]
\centerline{\epsfxsize=\figsize\epsffile{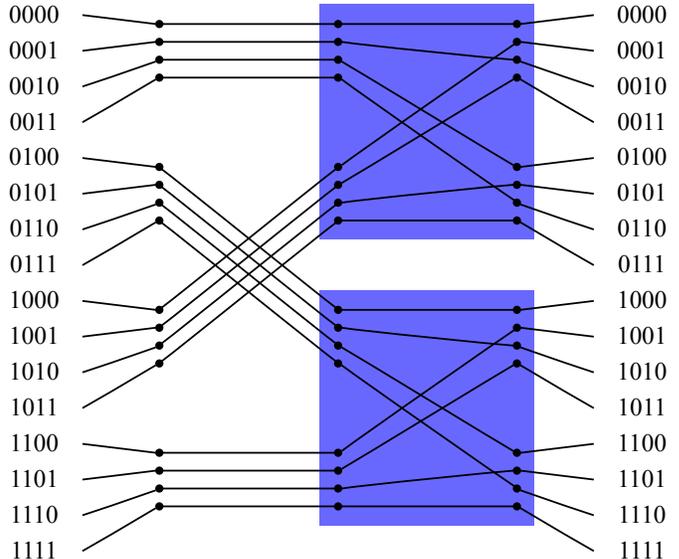}}
\caption{\label{16ShuffleFig}
A 16-link perfect shuffler built out of four cables, each carrying four links, and two eight-link perfect shufflers, shaded blue.
}
\end{figure}

If the left sides of $2^{n-k}$ cables, each carrying $2^k$ links were lined up such that the first cable (cable 0) carried links $0,\ldots,2^k-1$, the second carried links $2^k,\ldots,2\cdot2^k-1$, etc, then the leftmost $n-k$ bits of the link number could be circularly shifted one bit to the left by arranging the \emph{cables} into a perfect shuffle.  The rightmost $k+1$ bits could be circularly shifted one bit to the left by connecting each pair of adjacent cables (\ie all of the links sharing the leftmost $n-k-1$ bits after the first circular shift) into a board that had two cable inputs and two cable outputs, with the individual links in the cables arranged into a perfect shuffle.  This construction with $n=4$ and $k=2$ is shown in \fig{16ShuffleFig}.

From these two components, using $k=6$ (digital cables with 64 digital links are commercially available), each perfect shuffle layer in a $2^{20}$-link butterfly network would require 16,384 cables and 8192 two-cable perfect shufflers.  This is a nice saving over using single-link cables, especially in terms of the labor needed to assemble the layers.

By using larger circuit boards, many controlled swappers and many two-cable perfect shufflers could fit on a single circuit board, further cutting the number of parts.

As an additional improvement, the layers in a butterfly network can be rearranged---rather than using $n$ layers, each of which circularly shifts left by one bit and then xors the least significant bit, the network could instead use $n/\ell$ layers, each of which xors the least significant $\ell$ bits with an $\ell$-bit control and then circularly shifts by $\ell$ bits (if $n$ is not a multiple of $\ell$, then one layer would involve fewer bits).  A device that toggles the least significant $\ell$ bits of the address does not mix between blocks of $2^\ell$ bits, so a simple row of $2^\ell$-link $\ell$-bit togglers would make a $2^n$-link $\ell$ bit toggler.  This saves a factor of $\ell$ in the number of togglers needed; $\ell$ perfect shuffle layers could go between each toggler layer.

Putting this together for a $2^{20}$-link butterfly network, using $k=6$ and $\ell=8$ gives 20 perfect shuffle layers for a total of 327,680 cables and 163,840 two-cable perfect shufflers and three controlled xor layers containing a total of 12,288 256-link 8-bit togglers (of which one of the three layers would only use half of its control inputs).

Some further improvements would be possible by building layers that circularly shifted by more than one bit at a time---these could still use multiple-link cables but would need larger, although still probably inexpensive, multiple-bit perfect shuffle boards.

\subsection{Technology}
\label{TechnologySec}

So far, we have discussed layouts from an abstract point of view, without considering the particular technology used.  We will now briefly survey some attractive options for implementing this in practice.

\subsubsection{Off-the-shelf hardware}

The easiest design to prototype would use small (16-port, perhaps) nonblocking Ethernet switches and standard Ethernet cables, with one link per cable.  The switches are physically capable of performing any combination of swapping and shuffling on their ports, but they are not generally meant to change their connections on the fly.  Most so-called managed switches can, however, store a small number (often 4096) of fixed routes indexed by the destination of the packet that they are switching (this is called the forwarding table).  This means that, with some care, the destination field could be used to control the entire route taken by each packet traversing a butterfly network of Ethernet switches.  This could be inexpensive (a few tens of dollars per switch port in 2009 for one Gbps per link for a programmable managed switch and significantly more for 10 Gbps) but does not scale well beyond the square root of the size of the forwarding table, giving an upper bound of $N\le64$.  More flexible switches and routers are available, but tend to be far more expensive and slower.

FPGAs also allow rapid development.  Most FPGAs have both standard I/O pins, where a high voltage held for one clock cycle indicates a 1 bit and a low voltage indicates a 0 bit (these pins are limited to relatively low data rates on all but the most expensive FPGAs), and high-speed serial I/O pins, which can send or receive several gigabits per second on differential pairs of pins.  Any FPGA can easily act as an arbitrary shuffler or swapper, limited only by the numbers and types of I/O pins it has.  We experimented with FPGA implementations and found that a $2\times 2$ controlled swapper, even on a bottom-of-the-line Xilinx FPGA, could switch once per clock cycle, which was 64 million times per second on our device.  The downsides of FPGAs are their price and the fact that most FPGAs have relatively few high-speed I/O pins, keeping the cost per link quite high if data rates higher than one bit per clock per link are needed.

\subsubsection{Digital ASICs}

Custom application-specific integrated circuits (ASICs) can be fabricated in 2009 for a few hundred thousand dollars to make a set of masks plus very little for each part produced.  Any technology that can transmit and receive high-speed digital data can also be used to switch it.  For example, CMOS devices can switch at moderate speeds (several Gbps) and current-mode devices can operate in excess of 10Gbps per differential pair.  The number of links switched on a chip is limited only by the number of pins available on the chip, and a printed circuit board can hold as many of these chips as will fit at very low cost.

In fact, with a signaling technology that can tolerate enough loss, shufflers could be build on circuit boards containing no chips at all, keeping costs even lower.

\subsubsection{Analog switching}

Many protocols for transmitting large data rates over copper cables use advanced modulation techniques.  For example, gigabit Ethernet over CAT5 cabling uses multilevel signaling to achieve two gigabits per seconds (1Gbps each way) over four wire pairs at low frequency.  Devices to encode, decode, and error correct these kinds of protocols are complex and require significant power to operate, so it would be useful to minimize the number of times that data is modulated and demodulated in a butterfly network.  If we used analog switches that could exchange two \emph{modulated} signals with little enough loss, then we could have several layers of controlled swappers between each modulator/demodulator pair.

\subsubsection{Cable technologies}

Technologies to send large data rates over copper cable are well established.  Over short distances, a single conductor can carry one link.  Over longer distances, differential signals are usually sent over one pair of conductors per link, arranged into some form of transmission line, and different techniques can be used to modulate the signal depending on the frequency response of the transmission line, the available transmission power, and cost considerations.  Copper cables have the advantages of being relatively easy to construct, easy to connect, and they can interface easily with switching electronics.

Optical fibers, on the other hand, can carry much higher data rates than copper over a single fiber, and all-optical switching technologies (photonic or mechanical) can rapidly switch these high data rates with little loss.  Optical cables are inexpensive, but connecting them is labor-intensive and they are far more expensive to interface with electronics than copper.

Finally, it is possible to transmit very high data rates between boards without any cables at all using free-space optical communication, in which boards have lasers and photodiodes aimed at each other.  These laser beams can freely intersect each other, and devices that automatically aim and focus free-space optical links are available.  See \cite{all-optical} for an example of a large network switch built out of free-space optical links.  This technology could eliminate the need to hand-wire perfect shufflers altogether.

\section{Conclusions}
\label{ConclusionsSec}

The corner turner in a conventional interferometer ranges from moderately expensive (if the telescope is small enough to use an off-the-shelf Ethernet switch) to extremely expensive (if the number of antennas $N$ is very large).  
We have presented a corner turning algorithm based on a butterfly network topology which can solve the corner turning problem for any $N$ for an $\mathcal{O}(N \log N)$ hardware cost with a constant prefactor if 
inexpensive custom network parts are used.  Even for an interferometer with over one million antennas, the corner turner could require well under one million network parts, each of which could
cost only a few dollars in large volumes. By eliminating a key bottleneck, this bodes well for future large-$N$ radio telescopes.

\bigskip
{\bf Acknowledgements:}
The authors wish to thank Richard Bradley, Adrian Liu, Jacqueline Hewitt, Michael Matejek, Miguel Morales, Edward Morgan, Courtney Peterson, Scott Morrion, Joel Villasenor, Dan Werthimer, Chris Williams
and an anonymous referee for helpful comments. 
This work was supported by NASA grants NAG5-11099 and NNG 05G40G,
NSF grants AST-0607597, AST-0708534, AST-0907969, AST-0908848 and PHY-0855425, 
the DoD through the NDSEG Program,
and fellowships from the David and Lucile Packard Foundation.


\begin{thebibliography}{99}

\bibitem{SKAscienceBook}
\rfbook\nn Carilli C\dualand\nn Rawlings S eds.;2004;Science with the Square 
Kilometre Array;Elseview;{Amsterdam, \url{http://www.skads-eu.org/p/SKA_SciBook.php}}


\bibitem{Jester09}
\rf\nn Jester S\dualand\nn Falcke H;2009;New Astron.~Rev.;53;1

\bibitem{decadal1}
\rfprep\nn Furlanetto S {\etal};2009;{arXiv:0902.3011 [astro-ph]}

\bibitem{decadal2}
\rfprep\nn Furlanetto S {\etal};2009;{arXiv:0902.3259 [astro-ph]}

\bibitem{decadal3}
\rfprep\nn Komatsu E {\etal};2009;{arXiv:0902.4759 [astro-ph]}

\bibitem{decadal4}
\rfprep\nn Aguirre J {\etal};2009;{arXiv:0903.0902 [astro-ph]}

\bibitem{MoralesMoff}
\rfprep\nnn Morales M F;2008;{arXiv:0812.3669 [astro-ph]}

\bibitem{ATAMemoCornerTurn}
W. L. Urry, ``A Corner Turn Architecture.''  ATA Memo No. 14, 2000 Nov 17.

\bibitem{ATAMemoCornerTurn2}
W. L. Urry, M. Wright, M. Dexter \& D. MacMahn, ``The ATA Correlator", ATA memo 73.

\bibitem{ATAMemoNonsquare}
L. D'Addario, ``Generalization of the Memoryless Corner Turner to the
 Non-Square Case.''  ATA Memo No. 22, 2001 April 2.

\bibitem{fftt}
\rf\nn Tegmark M\dualand\nn Zaldarriaga M;2009;PRD;79;083530

\bibitem{omniscopes}
\rfprep\nn Tegmark M\dualand\nn Zaldarriaga M;2009;{arXiv:0909.0001 [astro-ph]}

\bibitem{CASPER}
\url{http://casper.berkeley.edu/}

\bibitem{Parsons08}
\rfprep\nn Parsons A;2008;{arXiv:0809.2266 [astro-ph]}

\bibitem{PAPER}
\rfprep\nnn Parsons A R;2009;arXiv:0904.2334

\bibitem{Youssef92}
\rf\nn Youssef A\dualand\nn Arden B;1992;Microprocess. Microsyst.;16;3

\bibitem{perfect-shuffle}
\rf\nnn Stone H S;1971;Computers, IEEE Transactions on;C-20;2

\bibitem{all-optical}
\rf\nn Hirabayashi K, \nn Yamamoto T, \nn Matsuo S\multiand\nn Hino S;1998;Appl. Opt.;37;2985



\end{thebibliography}
\end{document}